\documentclass[twocolumn,showpacs,aps,prc,superscriptaddress]{revtex4-1}

\usepackage{amssymb}
\usepackage[figuresright]{rotating}

\begin{document}

\title{Production of $e^{+}e^{-}$ from U+U collisions at $\sqrt{s_{NN}}$ = 193 GeV and Au+Au collisions at $\sqrt{s_{NN}}$ = 19.6, 27, 39, 62.4, and 200 GeV as measured by STAR}

\author{Joey Butterworth (for the STAR Collaboration)}
\affiliation{T.W. Bonner Nuclear Laboratory, Rice University, Houston, TX77005-1892, USA}

\begin{abstract}
  We present STAR's measurement of the $e^{+}e^{-}$ continuum as a function of centrality, invariant mass, and transverse momentum for U+U collisions at $\sqrt{s_{NN}}$ = 193 GeV.  Also reported are the acceptance-corrected $e^{+}e^{-}$ invariant mass spectra for minimum-bias Au+Au collisions at $\sqrt{s_{NN}}$ = 27, 39, and 62.4 GeV and U+U collisions at $\sqrt{s_{NN}}$ = 193 GeV.  The connection between the integrated $e^{+}e^{-}$  excess yields normalized by charge particle multiplicity ($dN_{ch}/dy$) at mid-rapidity and the lifetime of the fireball is discussed.
\end{abstract}

\maketitle

\section{Introduction}
\label{LIntro}
Relativistic heavy-ion collisions produced by the Relativistic Heavy Ion Collider (RHIC) are capable of generating a hot, dense, strongly interacting medium.  In order to study this medium, electromagnetic probes, such as $e^{+}e^{-}$ pairs, lend themselves as a natural choice.  The $e^{+}e^{-}$ are generated at all stages of the collision, interact electromagnetically, and thus, are able to traverse the strongly interacting medium with minimal effects on their final state while preserving information imprinted on them by their parent(s).
 Pairs with an invariant mass (M$_{ee}$) less than $\sim 1.2$ GeV/c$^{2}$ are of particular interest because it contains production from the $\rho$ meson.  
  The $\rho$ has been suggested to have its spectral function modified by the medium \cite{RappWamSpecFunc} and is in agreement with measurements reported in \cite{CERES,NA60,STAR200PRL,PHENIX,STAR19PLB,SHUAI}.  The in-medium modification of the $\rho$ spectral function is considered a possible link to chiral symmetry restoration \cite{CSR}.  Moreover, it has been suggested that the integrated yield of $e^{+}e^{-}$ production within the low mass region (LMR, M$_{ee}$ $\lesssim$ 1.2 GeV/c$^{2}$) can be related to the lifetime of the fireball \cite{RappLife}.  

The versatility of RHIC enables a systematic study of the $e^{+}e^{-}$ continuum.  The Beam Energy Scan Program has enabled the Solenoidal Tracker At RHIC (STAR) to collect enough statistics to study $e^{+}e^{-}$ production from Au+Au collisions at $\sqrt{s_{NN}}$ = 19.6, 27, 39, and 62.4 GeV while the total baryon density remained approximately constant, which the $\rho$ spectral function depends on.  
Here, the $e^{+}e^{-}$ continuum is studied as a function of $\sqrt{s_{NN}}$ and presented in the context of the fireball lifetime.  Additionally, RHIC's versatility is evident from the different colliding species that it can accelerate.  By switching from a Au+Au collision system at $\sqrt{s_{NN}}$ = 200 GeV to a U+U collision system at $\sqrt{s_{NN}}$ = 193 GeV, the collision energy density is expected in most central collisions to be up to 20\% higher than Au+Au collisions \cite{KIKOLA} while the collision energy remains within a couple percent.  If the energy density is increased, one may expect a longer fireball lifetime, and in turn, a larger $e^{+}e^{-}$ yield in the LMR.  Here, the $e^{+}e^{-}$ continuum from U+U collisions is studied and compared to the continuum from Au+Au collisions.

This paper presents a systematic study of the acceptance-corrected $e^{+}e^{-}$ production measured by STAR as a function of $\sqrt{s_{NN}}$ = 27, 39, and 62.4 GeV.  Furthermore, STAR measurements of the $e^{+}e^{-}$ continuum produced by U+U collisions are presented.

\begin{figure*}[t]
\begin{minipage}[b]{0.33\hsize}
\includegraphics[width=.99\textwidth]{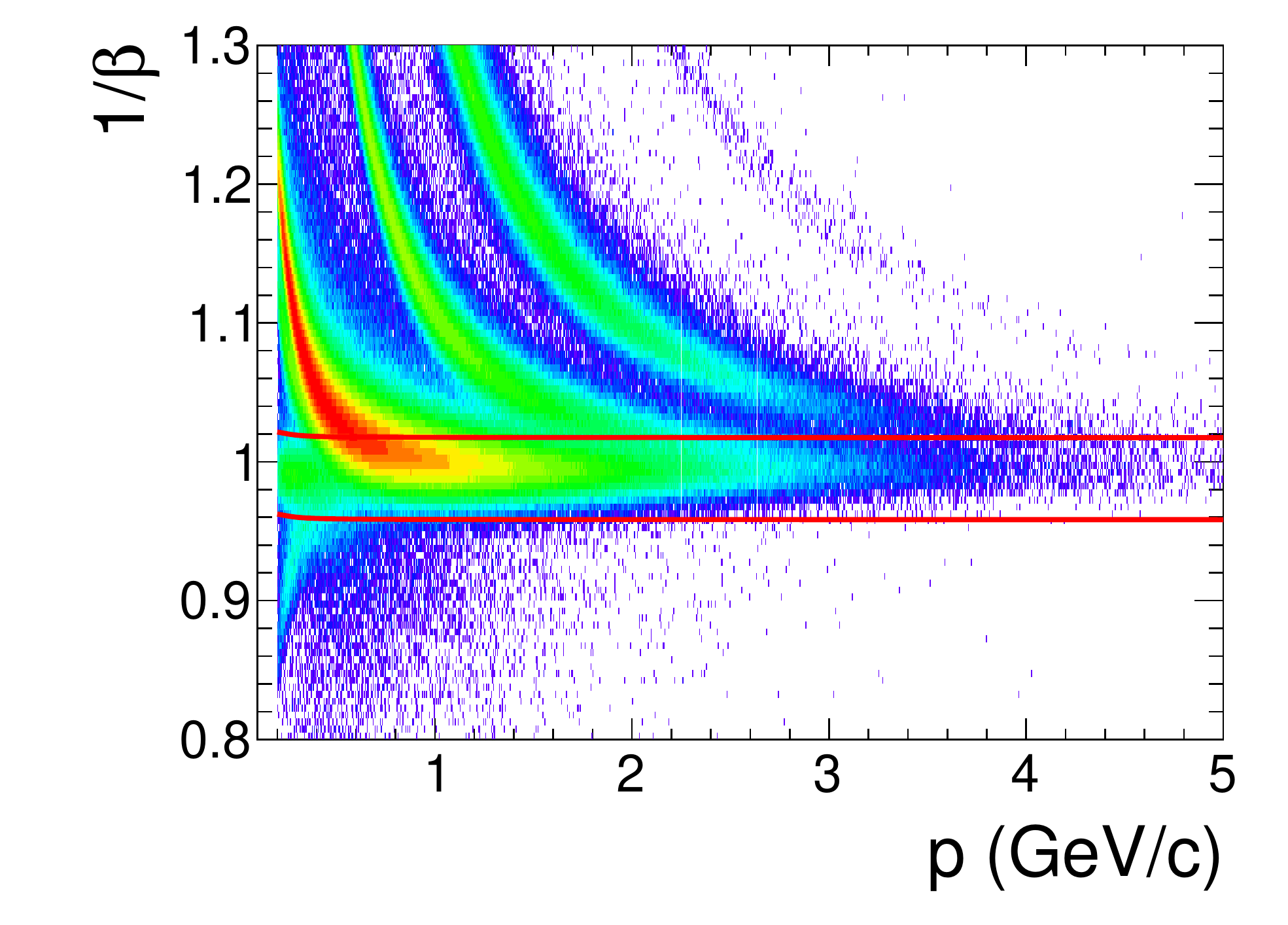}
\par\end{minipage}
\begin{minipage}[b]{0.33\hsize}
\includegraphics[width=.99\textwidth]{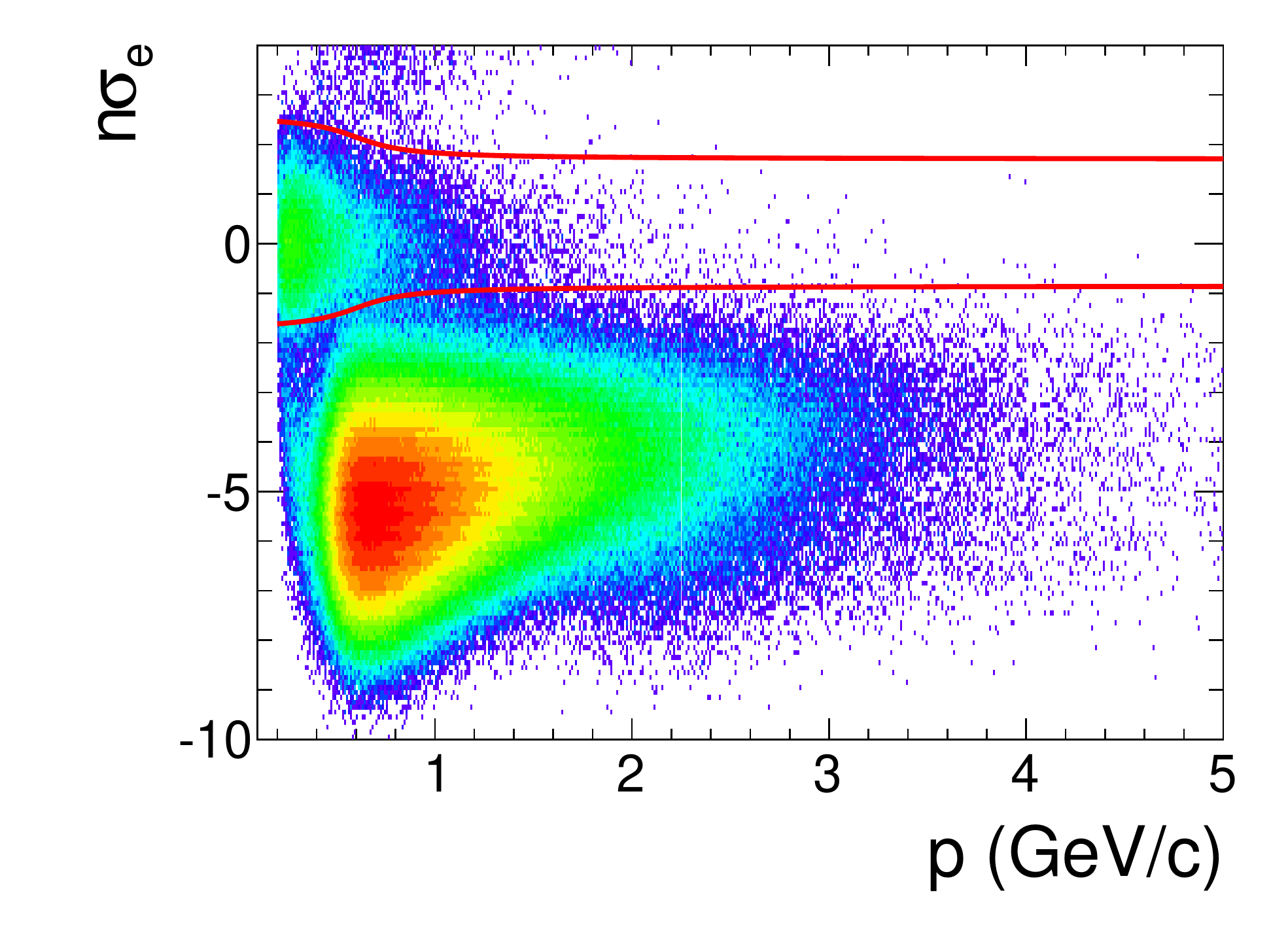}
\par\end{minipage}
\begin{minipage}[b]{0.33\hsize}
\includegraphics[width=.99\textwidth]{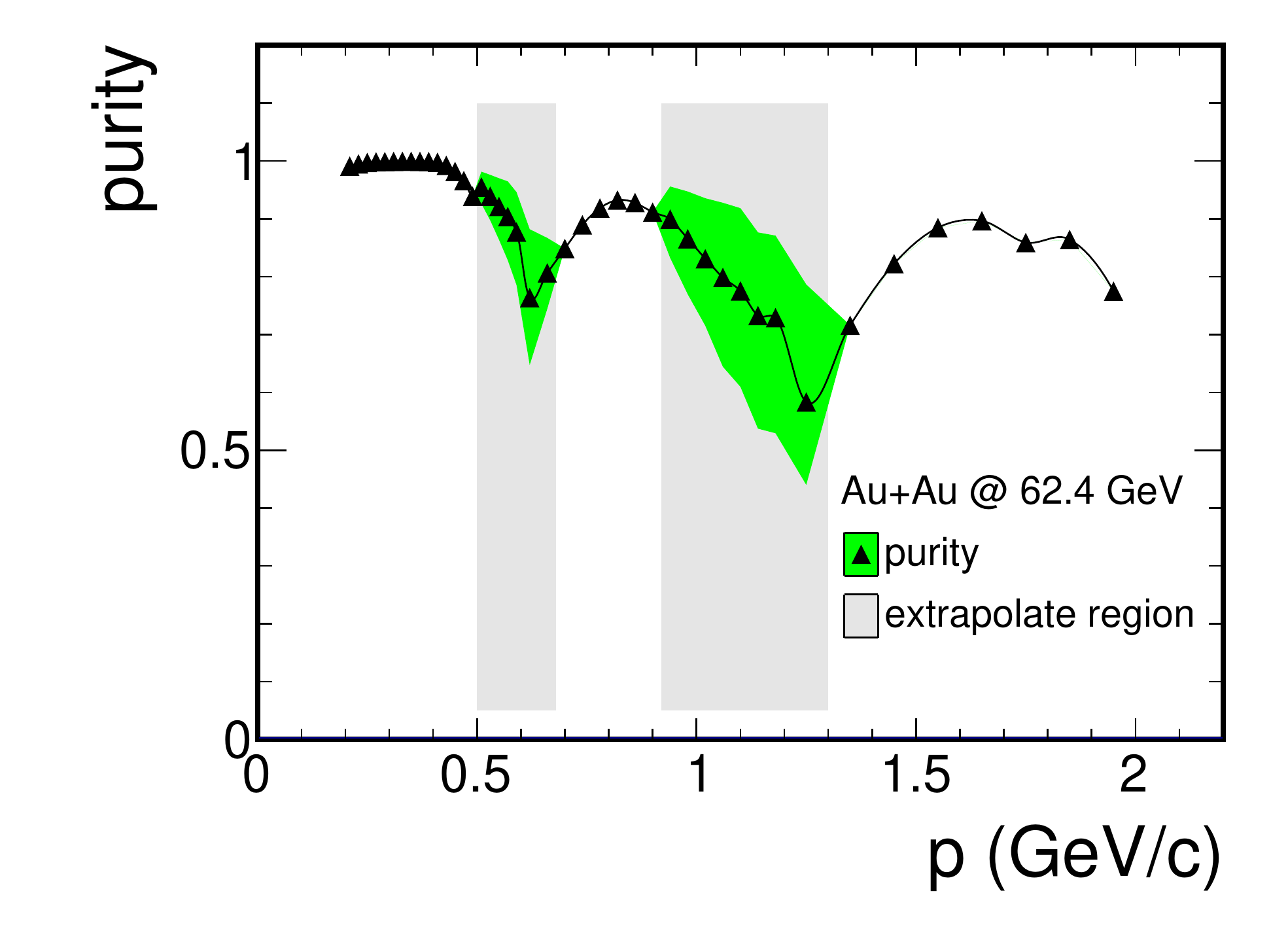}
\par\end{minipage}
\caption{(Color online)  (Left) $~\beta^{-1}$ as a function of momentum p for all tracks originating from the collision vertex.  The red lines depict the selection criteria used to reject slower hadrons while keeping electrons.  (Middle) The TPC's electron identification as a function of momentum after TOF velocity selection.  The red lines depict the selection criteria used to select a pure electron sample.  The bulb beneath the selected area are pions.  (Right)  The electron purity as a function of momentum where the green (online) bands represent systematic uncertainty and the solid gray region represents the region where the absolute yield of electrons and overlapping hadrons have been extrapolated.  }
\label{fig:62GeVeID}

\end{figure*}
\section{Data Sample and Analysis}
\label{LData}
STAR has collected 70M, 130M, and 67M events from the top-80\% most central Au+Au collisions at $\sqrt{s_{NN}}$ = 27, 39, and 62.4 GeV, respectively, and 270M events from the top-80\% most central U+U collisions at $\sqrt{s_{NN}}$ = 193 GeV, where the top-80\% represents STAR's minimum bias selection.  
The Time Projection Chamber (TPC) \cite{TPC} and Time of Flight (TOF) \cite{TOF} detectors are used to identify electrons and positrons within the STAR acceptance ($p_{T}^{e}$ $>$ 0.2 GeV/c,  $|\eta^{e}|$ $<$ 1, and $|y_{ee}|$ $<$ 1).  The TPC provides the particle identification and tracking via ionization energy loss (dE/dx) while the TOF is used to reject the slower hadrons then enhancing the TPC's particle identification and purity.  This is depicted in Fig.\ \ref{fig:62GeVeID}, where the left panel demonstrates the TOF velocity selection criteria (between the red lines), the middle panel shows the electron sample selection based on the TPC electron identification (n$\sigma_{e}$) after the TOF velocity rejection of slower hadrons.  The right panel illustrates the purity of the selected electrons as a function of momentum, the average purity for each data sample is at the level of 95\% or greater.

\begin{figure}[!htb]
\begin{minipage}[b]{0.95\hsize}
\centering

\includegraphics[width=1.\textwidth]{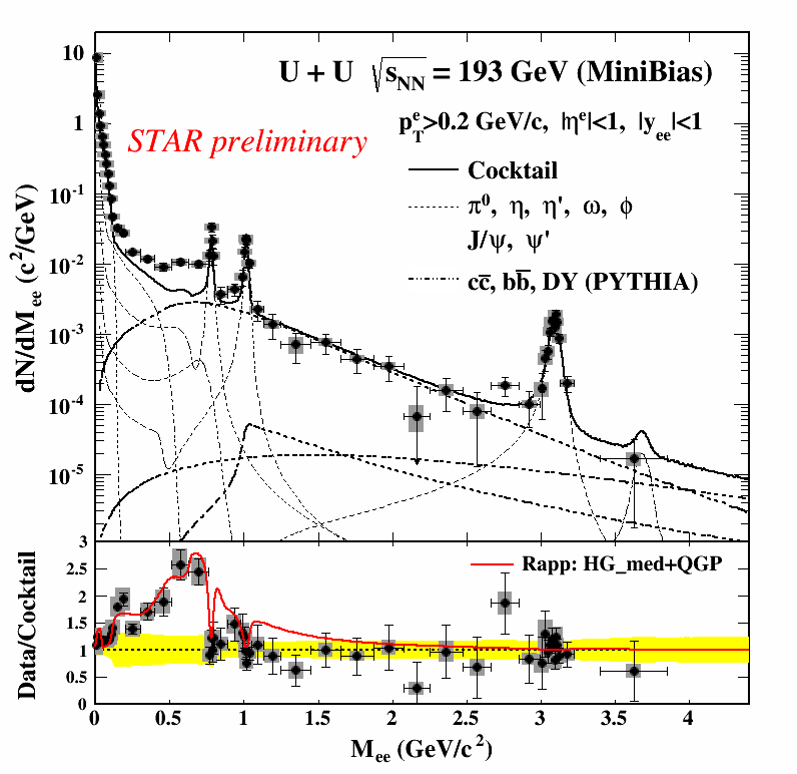}

\caption{(Color online) (Top) The corrected $e^{+}e^{-}$ invariant mass spectrum from U+U collisions (black markers).  The perforated lines represent known hadronic contributions and the cocktail (solid line) is a summation of these contributions.  (Bottom) The data over cocktail ratio as a function of invariant mass.  The red line is the ratio of Rapp \emph{et al.}\ \cite{RappWamSpecFunc,RappPrivate} calculations plus the cocktail to the cocktail. The systematic uncertainties are shown as the shaded areas and the statistical uncertainties as the error bars.  The yellow region along the dotted line at data/cocktail = 1 represents the cocktail uncertainty. }
\label{fig:UUmb_ratio}
\end{minipage}
\end{figure}

Selected electrons (and positrons) are combined to form the $e^{+}e^{-}$ foreground.  This contains background from the combinatorial background and correlated pairs (e.g. jets and double Dalitz decays).  To estimate and remove the background, a geometric mean with a charge-acceptance correction is subtracted from the uncorrected $e^{+}e^{-}$ distribution as a function of the pair invariant mass and pair transverse momentum ($p_{T}^{ee}$). 
After subtraction, the continuum is then corrected for efficiency and acceptance losses.  Details on the methods used may be found here \cite{STAR200PRC}.  Shown in the top panel of Fig.\ \ref{fig:UUmb_ratio} is the $e^{+}e^{-}$ continuum as a function of invariant mass for minimum bias U+U collisions at $\sqrt{s_{NN}}$ = 193 GeV.  Known hadronic contributions to the invariant mass spectrum are then modeled and compared to the data.  Contributions are modeled from $\pi^{0}$, $\eta$, $\eta$', $\omega$, $\phi$, J/$\psi$, $c\bar{c}$, Drell-Yan, and $b\bar{b}$, where Drell-Yan and $b\bar{b}$ are only modeled for U+U collisions at $\sqrt{s_{NN}}$ = 193 GeV and Au+Au collisions at $\sqrt{s_{NN}}$ = 200 GeV.  The top panel in Fig.\ \ref{fig:UUmb_ratio} also shows the known hadronic contributions as perforated lines and the cocktail sum of their contributions shown as the solid line.

\begin{figure}[!htb]
\begin{minipage}[b]{0.95\hsize}
\centering
\includegraphics[width=1.\textwidth]{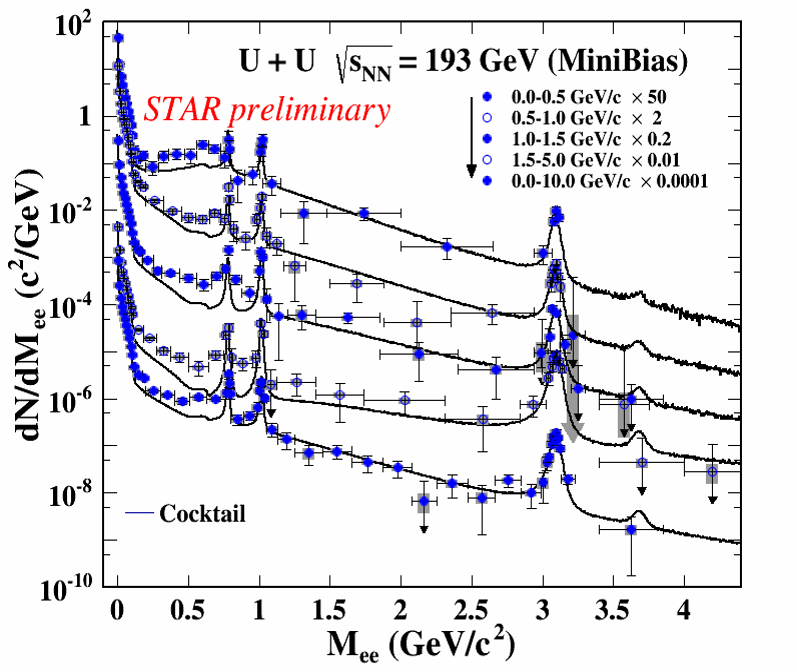}
\caption{(Color online) The $e^{+}e^{-}$ invariant mass spectrum (markers) for different $p_{T}^{ee}$ ranges.  The systematic uncertainties are represented by the shaded regions and the statistical uncertainties are represented by the error bars.  The hadronic cocktail is shown for each $p_{T}^{ee}$ range (solid line).}
\label{fig:UUpt}
\end{minipage}
\end{figure}

\begin{figure}[!htb]
\begin{minipage}[b]{0.95\hsize}
\centering
\includegraphics[width=1.\textwidth]{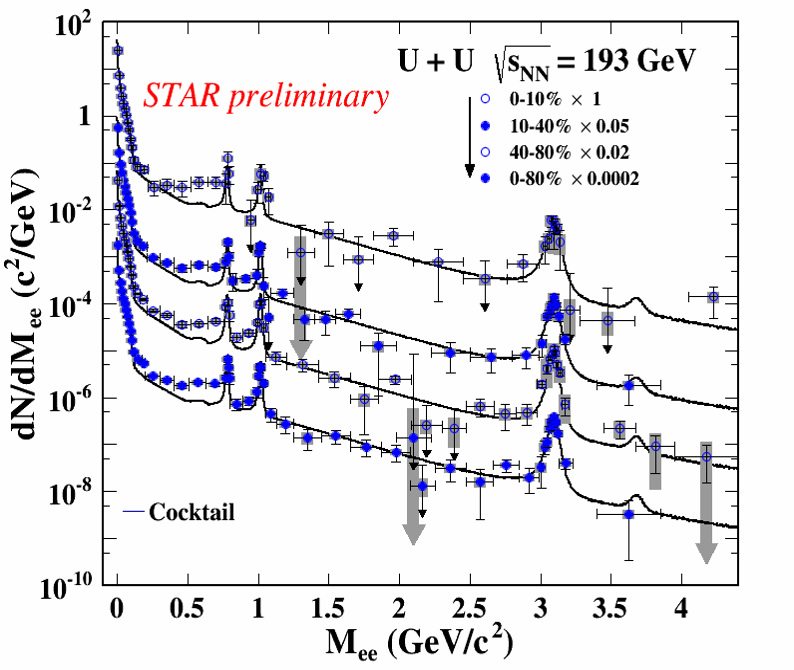}
\caption{(Color online) The $e^{+}e^{-}$ invariant mass spectrum (markers) for different centrality ranges.  The systematic uncertainties are represented by the shaded regions and the statistical uncertainties are represented by the error bars.  The hadronic cocktail is shown for each centrality range (solid line).}
\label{fig:UUCent}
\end{minipage}
\end{figure}

Figures \ref{fig:UUpt} and \ref{fig:UUCent} are the $e^{+}e^{-}$ invariant mass spectrum for different $p_{T}^{ee}$ and centrality ranges, respectively.  For reference, the all-inclusive distribution from Fig.\ \ref{fig:UUmb_ratio} is shown as the bottom distribution in each figure.

\section{Results and Discussion}
\label{LResults}
\begin{figure}[!htb]
\begin{minipage}[b]{0.95\hsize}
\centering

\includegraphics[width=1.\textwidth]{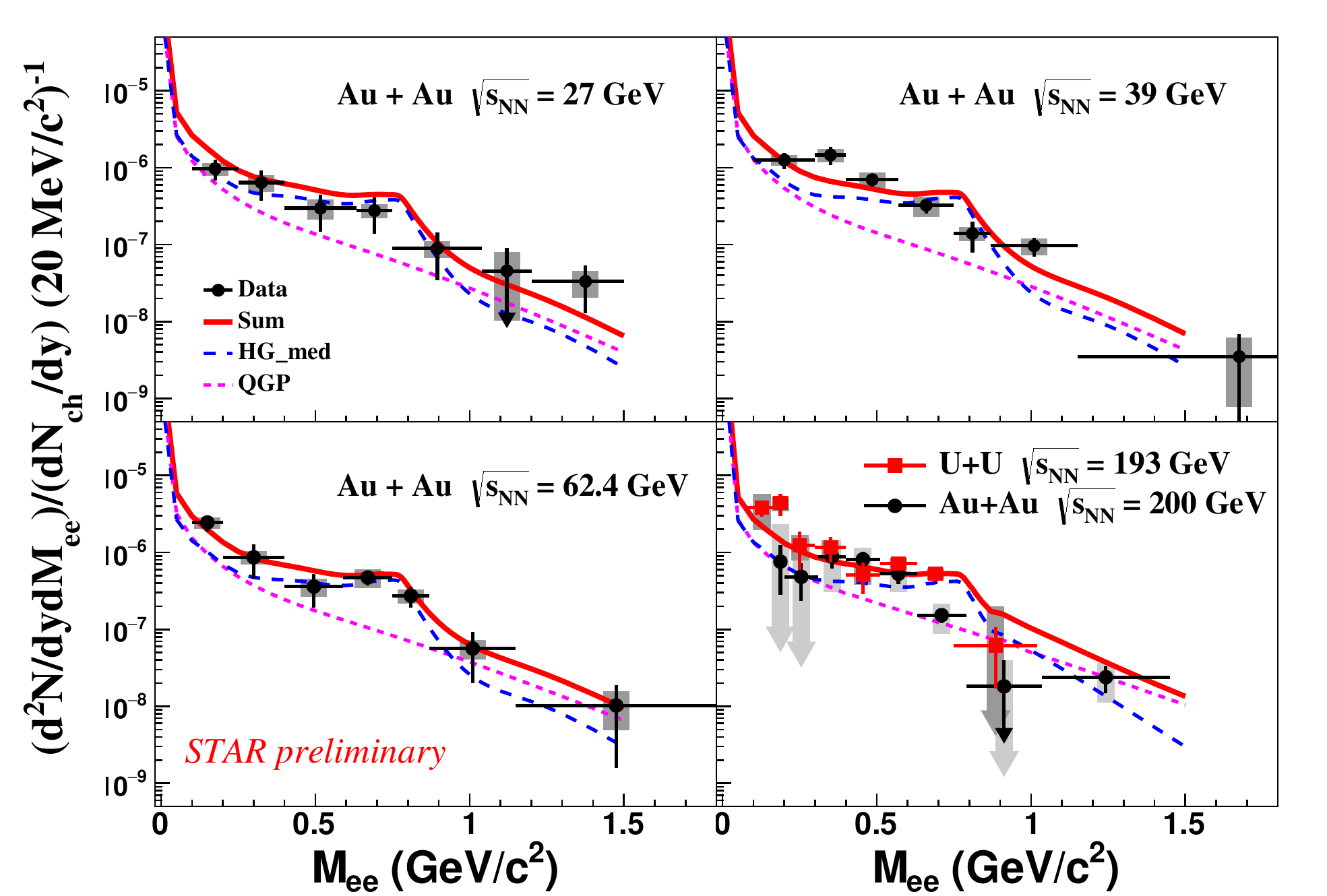}

\caption{(Color online) The acceptance-corrected $e^{+}e^{-}$ invariant mass excess yield normalized by $dN_{ch}/dy$.  Data from minimum bias Au+Au collisions (black markers) are shown in all four panels at $\sqrt{s_{NN}}$ = 27, 39, 62.4, and 200 GeV, starting from the top and reading from left to right, respectively.  Data from minimum bias U+U collisions (red markers) at $\sqrt{s_{NN}}$ = 193 GeV is shown in the lower right panel.  Systematic uncertainties are the shaded areas and statistical uncertainties are the error bars.  Calculations from Rapp \emph{et al.}\ \cite{RappWamSpecFunc,RappPrivate} are shown in each panel for contributions from the hadronic medium (dashed blue curve), QGP (dashed pink curve), and their sum (solid red curve).  In the lower right panel, the Rapp \emph{et al.}\ calculations are for U+U collisions.}
\label{fig:AccCorrBESUU}
\end{minipage}
\end{figure}

  Figures \ref{fig:UUmb_ratio}, \ref{fig:UUpt}, and \ref{fig:UUCent} exhibit a difference between the invariant mass spectrum and the hadronic cocktail without $\rho$ contributions, or excess.  A model by Rapp \emph{et al.}\ that incorporates broadening of the $\rho$ spectral function (HG\_med) and QGP thermal radiation is in agreement with this observation.  This is demonstrated by overlaying the ratio of the model calculations plus the cocktail to the cocktail and comparing to the data over cocktail ratio in the bottom panel of Fig.\ \ref{fig:UUmb_ratio}.  This has previously been shown for Au+Au collisions at $\sqrt{s_{NN}}$ = 19.6, 27, 39, 62.4, and 200 GeV \cite{STAR200PRL,STAR19PLB,Patrick}.  Taking a step further, the excess is corrected for STAR's kinematic acceptance.  The acceptance-corrected excess at mid-rapidity has been normalized by the charge particle multiplicity at mid-rapidity ($dN_{ch}/dy$) to cancel volume effects and is presented in Fig.\ \ref{fig:AccCorrBESUU} for U+U collisions at $\sqrt{s_{NN}}$ = 193 GeV and Au+Au collisions at $\sqrt{s_{NN}}$ = 27, 39, 62.4, and 200 GeV.  In the same figure, we compare our results with model calculations from Rapp \emph{et al.}\ that incorporate thermal radiation from the QGP and a broadened $\rho$ spectral function.  

\begin{figure}[!htb]
\begin{minipage}[b]{0.95\hsize}
\centering

\includegraphics[width=1.\textwidth]{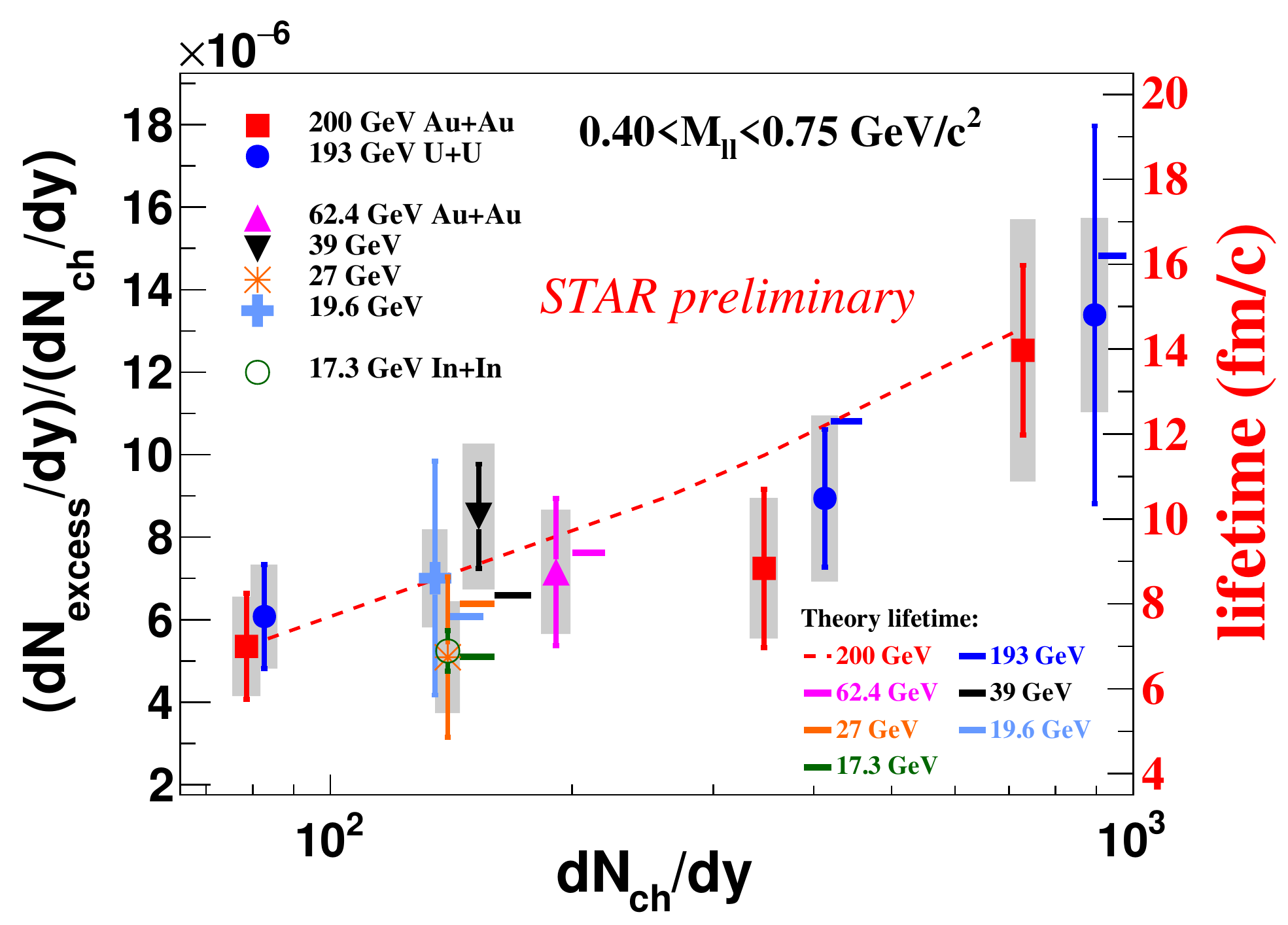}

\caption{(Color online) (Left y-axis) The integrated acceptance-corrected $e^{+}e^{-}$ excess yield normalized by $dN_{ch}/dy$ as a function of $dN_{ch}/dy$, where data markers represent each point. Statistical uncertainties are represented by the error bars and the systematic uncertainties are represented by the shaded regions.  (Right y-axis) The lifetime of the fireball calculations by Rapp \emph{et al.}\ \cite{RappLife,RappPrivate} as a function of $dN_{ch}/dy$, where the calculations are represented by the solid bars that have been offset in the +x-direction and the red dashed curve are lifetime calculations for Au+Au at $\sqrt{s_{NN}}$ = 200 GeV.   }
\label{fig:IntExcess}
\end{minipage}
\end{figure}

To study the possible connection between the lifetime and excess yield, the $e^{+}e^{-}$ excess yield has been integrated from 0.4 to 0.75 GeV/c$^{2}$, normalized by $dN_{ch}/dy$, and plotted in Fig.\ \ref{fig:IntExcess} as a function of $dN_{ch}/dy$.  On the same figure, lifetime calculations from Rapp \emph{et al.}\ \cite{RappLife,RappPrivate} are plotted for each corresponding yield measurement as a bar while the trend for Au+Au at $\sqrt{s_{NN}}$ = 200 GeV centrality calculations is plotted as a dashed line.  There is an increase in normalized yields at higher collision energies with respect to the lower energies, and at $\sqrt{s_{NN}}$ = 200 GeV there is an increase in normalized yields in more central collisions with respect to peripheral collisions. The expected lifetime in model calculations also has an increasing trend from 
peripheral to central collisions \cite{STAR19PLB}.

\section{Summary and Outlook}
\label{LSummary}
  We have presented STAR measurements of the $e^{+}e^{-}$ invariant mass for U+U collisions at $\sqrt{s_{NN}}$ = 193 GeV along with the acceptance-corrected excess $e^{+}e^{-}$ yields for minimum bias U+U collisions at $\sqrt{s_{NN}}$ = 193 GeV and minimum bias Au+Au collisions at $\sqrt{s_{NN}}$ = 27, 39, 62.4, and 200 GeV.  Comparisons between these measurements and model calculations, which include broadening of the $\rho$ spectral function and QGP thermal radiation, have been shown to be in agreement.  Also, we reported the normalized integrated excess yields as a function of $dN_{ch}/dy$ for various collision systems.  These measurements are consistent with theoretical calculations that indicate a longer fireball lifetime for collisions that are central and have a higher $\sqrt{s_{NN}}$.
  
  Measurements made during the Beam Energy Scan Program were made at approximately the same total baryon density.  At $\sqrt{s_{NN}}$ $<$ 20 GeV, the total baryon density rises as $\sqrt{s_{NN}}$ decreases.  Models such as \cite{RappWamSpecFunc} suggest that the $\rho$ spectral function is dependent upon the total baryon density; hence, if the total baryon density increases, the $\rho$ yield is expected to increase too.  To test this relation and distinguish between models, STAR will take advantage of RHIC's second Beam Energy Scan Program \cite{BESIIWP} where Au+Au will be collided at $\sqrt{s_{NN}}$ = 7.7, 9.1, 11.5, 14.5, and 19.6 GeV.  It is expected that the number of events recorded will provide statistical uncertainties similar to the statistical uncertainties quoted in STAR's Au+Au measurement at $\sqrt{s_{NN}}$ = 200 GeV \cite{STAR200PRL}.  In addition to the increased statistics, the inner sector of the TPC will be replaced to provide additional tracking points \cite{iTPC} and a complimentary forward Time of Flight (eTOF) detector will be installed \cite{eTOF}.  The TPC upgrade will lead to additional reduction in the statistical and systematic uncertainties, while the eTOF will lead to an extension in the rapidity reach and the ability to measure the $e^{+}e^{-}$ dependence on rapidity.

\end{document}